\documentclass[3p,10pt]{elsarticle}

\usepackage{amssymb}

\usepackage{graphicx}
\usepackage{balance}
\usepackage{color}
\usepackage{xcolor}
\usepackage{multirow}
\usepackage{caption}

\usepackage{subfigure,epsfig,amsfonts,amsmath,cases,amssymb,graphics, latexsym, times, algorithmic, algorithm, bbm, color, soul,wrapfig,comment}

\usepackage{makeidx,epsf}
\usepackage{epsfig}

\newproof{proof}{\bf Proof}

\usepackage{wrapfig}

\journal{Neural Network}

\begin{document}

\begin{frontmatter}



\title{Adaptive Generation of Phantom Limbs Using Visible Hierarchical Autoencoders}


\author[utc-cs] {Dakila Ledesma}

\author[utc-cs]{Yu Liang*}
\ead{yu-liang@utc.edu}

\author[utc-cs] {Dalei Wu}

\address[utc-cs]{Deparment of Computer Science and Engineering, University of Tennessee at Chattanooga, United States}

\cortext[cor1]{Corresponding author}

\begin{abstract}

This paper proposed a hierarchical visible autoencoder in the adaptive phantom limbs generation according to the kinetic behavior of functional body-parts, which are measured by heterogeneous kinetic sensors. The proposed visible hierarchical autoencoder consists of interpretable and multi-correlated autoencoder pipelines, which is directly derived from hierarchical network described in forest data-structure. According to specified kinetic script (e.g., dancing, running, etc.) and users' physical conditions, hierarchical network is extracted from human musculoskeletal network, which is fabricated by multiple body components (e.g., muscle, bone, and joints,  etc.) that are bio-mechanically, functionally, or nervously correlated with each other and  exhibit mostly non-divergent kinetic behaviors. Multi-layer perceptron (MLP) regressor models as well as several variations of autoencoder models are investigated for the sequential generation of missing or dysfunctional limbs. The resulting kinematic behavior of phantom limbs will be constructed using virtual reality and augmented reality (VR/AR), actuators, and potentially controller for prosthesis (an artificial device that replaces a missing body part).  The addressed work aims to develop practical innovative exercise methods that (1) engage individuals at all ages, including those with chronic health condition(s) and/or disability, in regular physical activities, (2) accelerate the rehabilitation of patients, and  (3) release users' phantom limb pain. The physiological and psychological impact of the addressed work will critically assessed in future work. 
\end{abstract}

\begin{keyword}
	Visible autoencoder, human musculoskeletal network, phantom limb,  virtual reality, kinetics.
\end{keyword}

\end{frontmatter}

	\section{Introduction}
	\label{Section:instruction}

This paper proposed a hierarchical auto-encoder neural network \cite{Goodfellow2016} in the adaptive phantom limbs generation (denoted as PLG-HAE for simplicity) according to the kinetic behavior of functional body-parts, which are measured by variance of kinetic sensors. Clinically, a phantom limb \cite{Ortiz-2016-Phantom-Pain} is defined by the sensation that an amputated or missing limb is still attached \cite{Ramachandran1998-PhantomLimb}. To counteract phantom limbs, this work employs sensors and actuators, virtual reality and augment reality, and deep learning to generate missing or dysfunctional limbs. As a an intelligent, four-dimensional (namely X-Y-Z plus somatosensory), partial control (e.g., phantom limb generation), and virtual-reality-enabled rehabilitation modality, this project directly influences the user's motivation for movement \cite{Liang2018-Elsevier-SH}. 

%
\begin{figure}[htp]
 	\centering
	\includegraphics[height=3.5 cm]{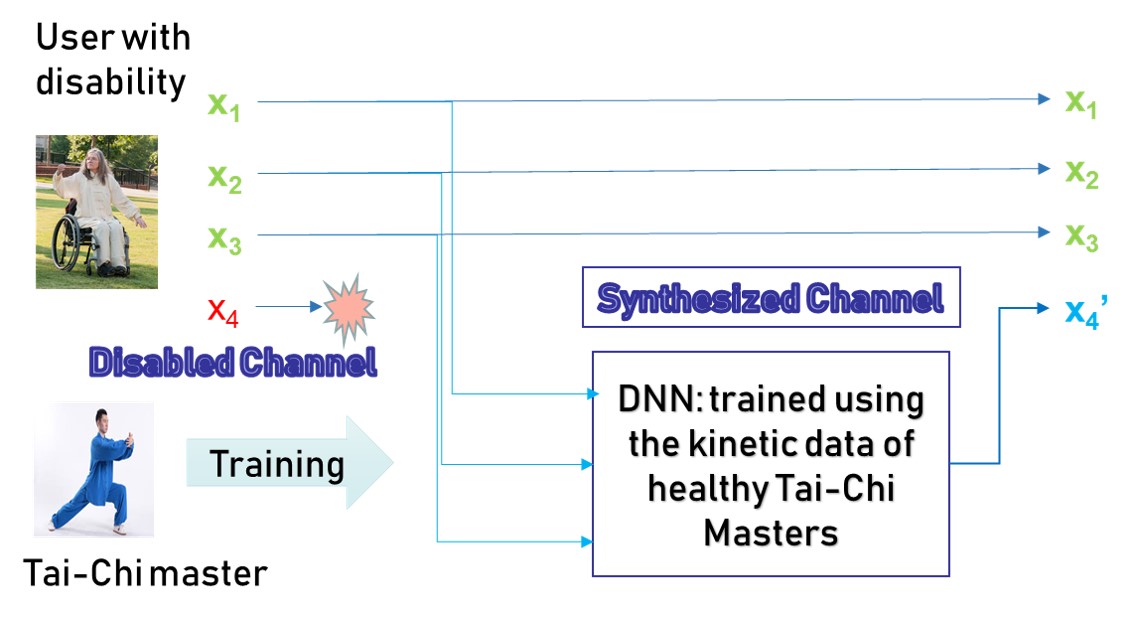}
	\includegraphics[height=3.5 cm]{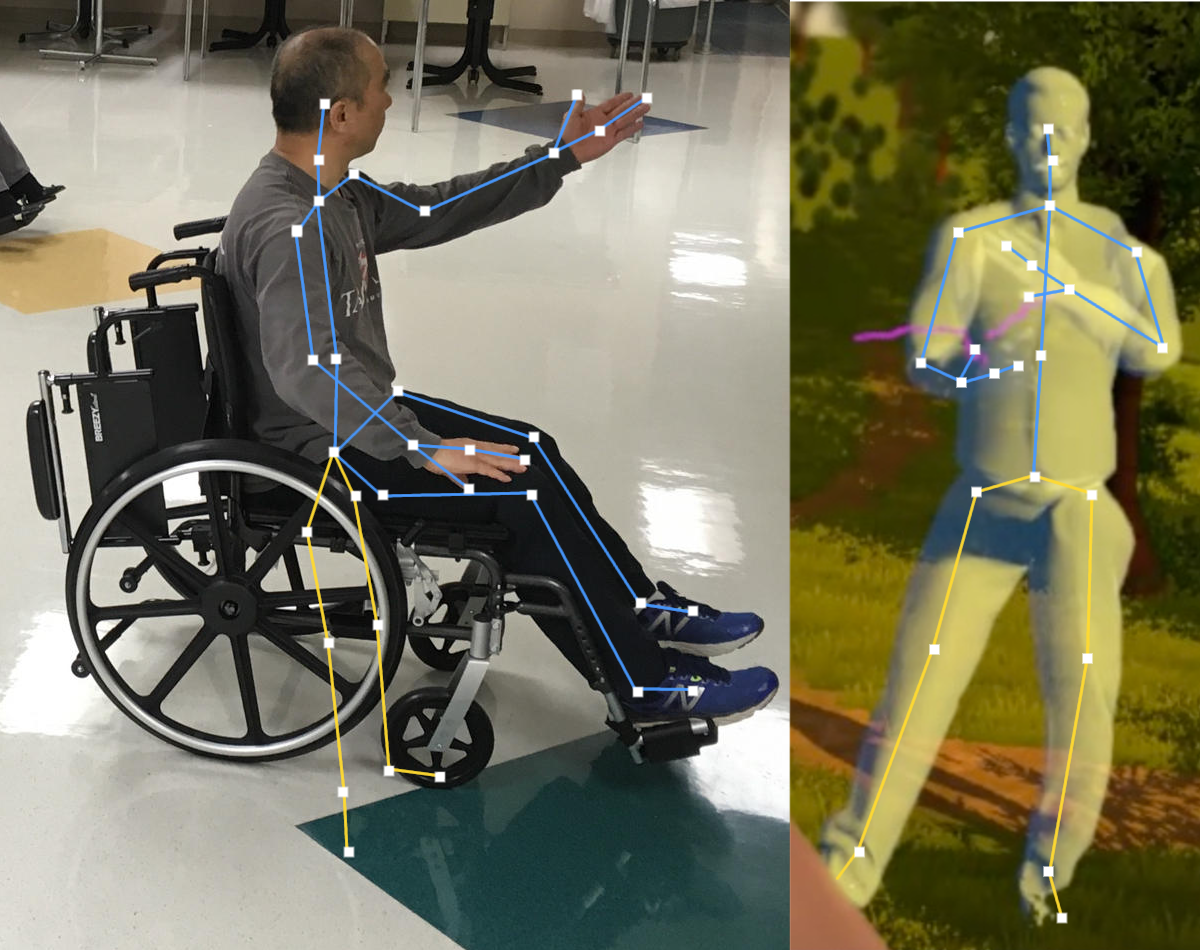}
	\includegraphics[height=3.5 cm]{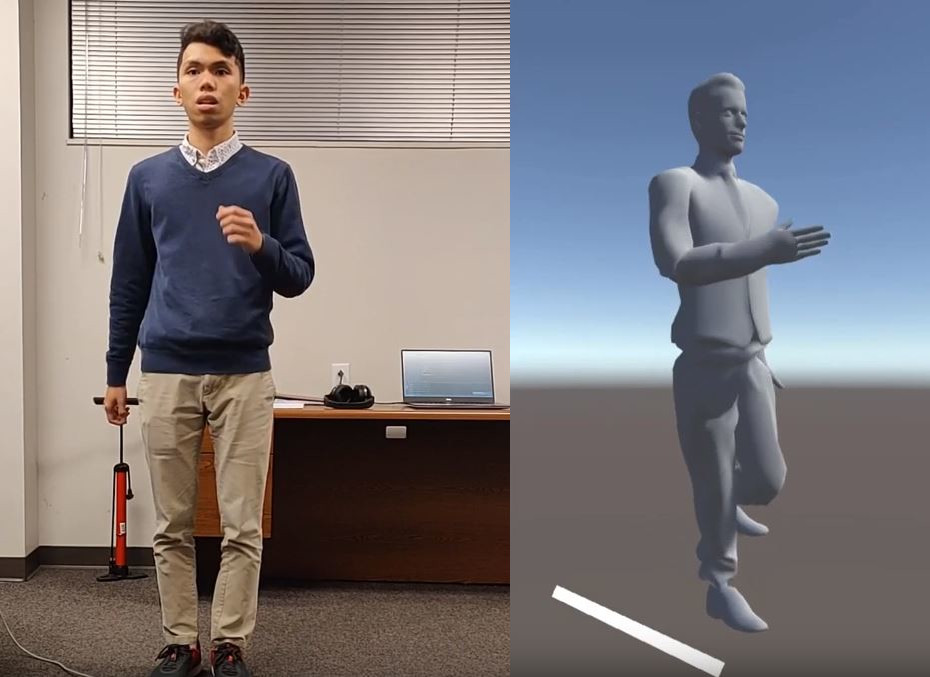}
	\caption{Compensation of disabled input channels via Deep Neural Network: (a) framework; (b) snapshot about the compensated kinetic status of a wheel-chaired user, who receives ``functional phantom legs'' (in yellow color) through disable-channel compensation;  (c) walking by moving arms (online video \cite{online13-Dakila-Autoencoder-PhantomLeg2018}) 
	}
	\label{Fig16-CompensateDisabledChannels}
\end{figure} 

It is known that various components of human body are correlated functionally, bio-mechanically, or nervously. As a machine learning strategy, the proposed work generates phantom limbs from other functional and measurable body parts by employing above correlations among human body \cite{Ortiz-2016-Phantom-Pain}.
As illustrated in Figure \ref{Fig16-CompensateDisabledChannels}(a), $x_1$, $x_2$, $x_3$, and $x_4$ constitute the four input channels while the lower-limb-related input $x_4$ is not valid for a seated user. Fortunately, we can formulate $x_4$ according to the correlation between $x_4$ and other enabled inputs -- $x_1$, $x_2$, and $x_3$. The correlation can be discovered using a supervised deep neural network (DNN) regression \cite{Bengio2009,Sperduti1997}, based on which the input $x_4$ can be reconstructed as a function of other known inputs: $x_4' = DNN(x_1,x_2,x_3) $. Tai-Chi movement is selected as the script, there the training data comes from the recorded virtual Tai-Chi master's kinetic data \cite{Liang2018-CHASE, Liang2018-Elsevier-SH}. As our preliminary contribution, MLP and autoencoder methods are employed to construct phantom legs according to arm kinetics.

Figure \ref{Fig16-CompensateDisabledChannels}(b) shows a snapshot about the compensated kinetic status of a wheelchair-bound user, who received ``functional legs,'' in compensation for the phantom limbs through compensation of missing data. In this figure, the compensation of missing data is generated by a neural network, and the visualization of this generated data is done through a virtual environment, namely VR. Besides bringing to the forefront improvements to body gesture recognition and motion generation, missing data compensation has potential clinical benefit to users, i.e. as a treatment to phantom limb pain \cite{Ortiz-2016-Phantom-Pain}. 

\begin{figure}[htp]
 	\centering
	\includegraphics[height=3.8 cm]{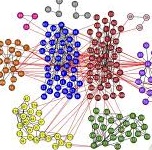}
	\hspace{0.4cm}
	\includegraphics[height=3.8 cm]{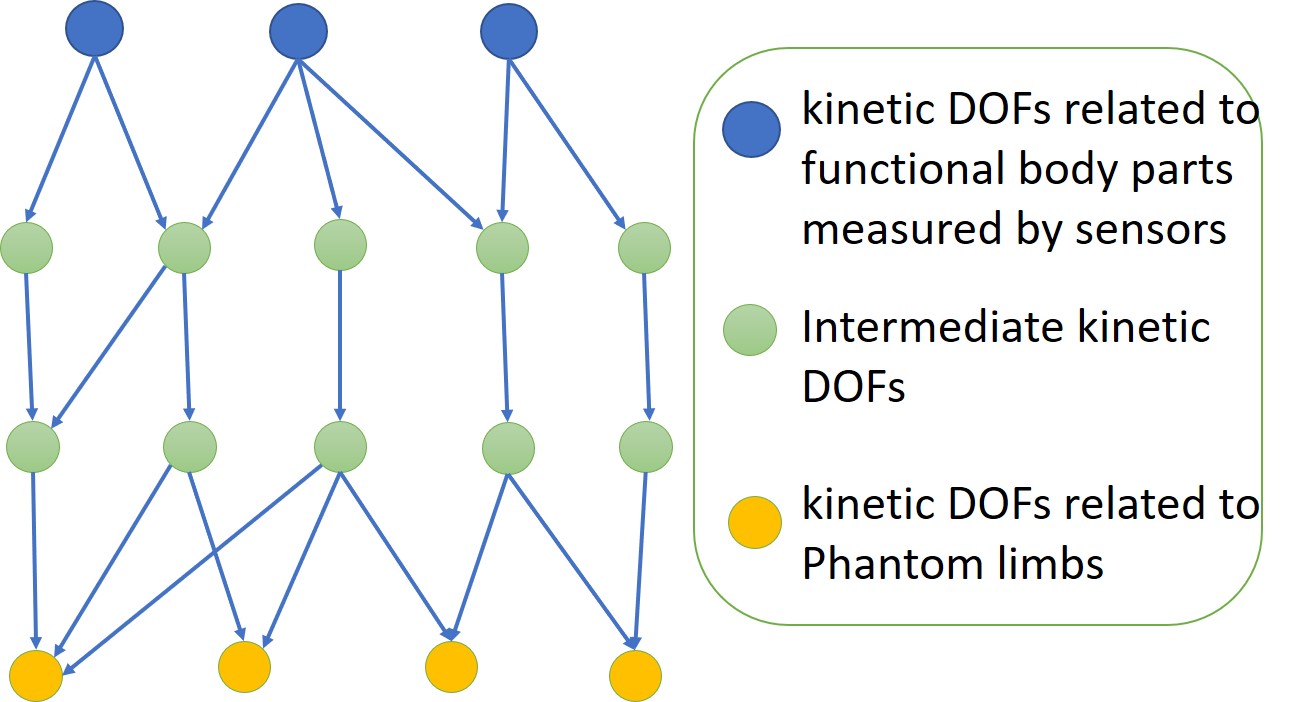}
	\hspace{0.4cm}
	\includegraphics[height=3.8 cm]{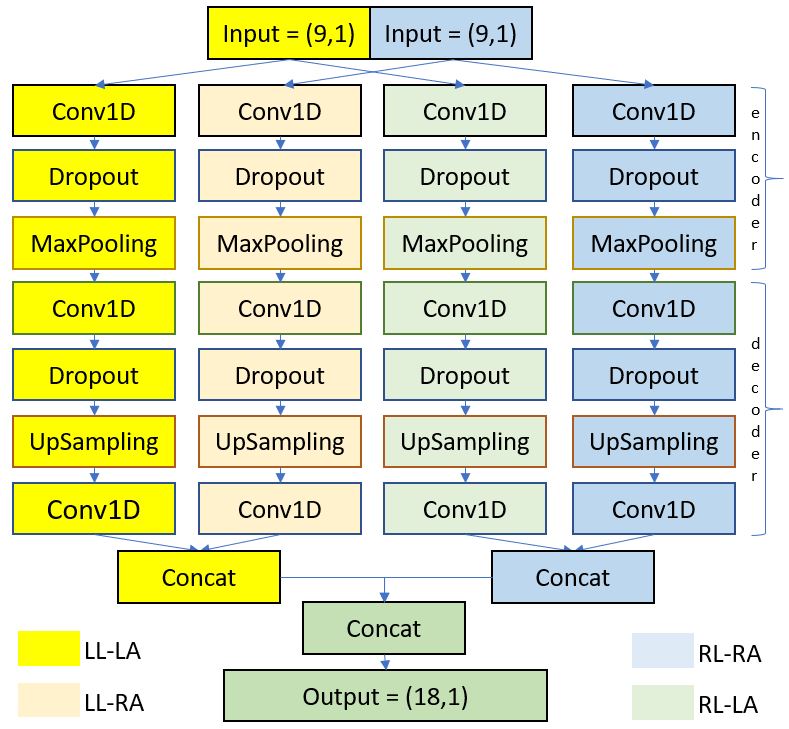}
	\caption{Derivation  of visible autoencoder Neural Network: (a) muscoloskeletal network; (b) sample hierarchical network in forest data structure;  (c) visible autoencoder neural network derived from hierarchical network.
	}
	\label{Fig01-derivation-visible-autoencoder}
\end{figure} 

As illustrated in Figure \ref{Fig01-derivation-visible-autoencoder}, the proposed visible hierchical autoencoder consists of interpretable and multi-correlated autoencoder pipelines, which is directly derived from hierarchical network described in forest data-structure. According to specified kinetic script (e.g., dancing, running, etc.) and users' physical conditions, hierarchical network is extracted from human musculoskeletal network, which is fabricated by multiple body components (e.g., muscle, bone, and joints,  etc.) that are bio-mechanically, functionally, or nervously correlated with each other and  exhibit mostly non-divergent kinetic behaviors. 

For the purposes of the neural network, the motion similarity between these correlations are assessed to improve performance. Multi-layer perceptron (MLP) regressor models, as well as several variations of autoencoder models are investigated for the sequential compensation of missing or dysfunctional limbs. The resulting kinematic behavior of phantom limbs will be constructed using virtual reality and augmented reality (VR/AR), actuator, and potentially controller for prosthesis (an artificial device that replaces a missing body part).  

The addressed work aims to develop practical innovative exercise
methods that (1) engage individuals at all ages, including those with chronic health condition(s) and/or disability, in
regular physical activities, (2) accelerate the rehabilitation of patients, and  (3) release the phantom limb pain.   
The physiological and psychological impact of the addressed work will critically assessed in future work. 

The remainder of this paper is organized as follows. Section 2 overviews the implementation of the generation system. Section 3 explains the data acquisition, transmission, and visualization of data. Section 4 investigates the human musculoskeletal network and some of its correlations to the intuitions used for the proposed neural network. Section 5 explains the data preprocessing methods for the autoencoder. Section 6 explains and reports the neural network architecture. Section 7 reports the the results of using our proposed network. Lastly, Sections 8 and 9 concludes the paper and defines future work.

\section{Implementation of System}
\label{sec:implementation-system}

The proposed work employs deep learning techniques such as autoencoder to generate phantom limbs according to the observed kinetic behaviors of other body parts based on the following hypothesis:
(1) The human body consists of multiple components such as muscle, bones, and joints, which are correlated with each other mechanically, nervously, or functionally. 
(2) Because deep learning techniques such as autoencoder can be used to formulate the pattern of specific kinetic behavior script such as dancing and sports.

\begin{figure}[htp]
 	\centering
	\includegraphics[height=4 cm]{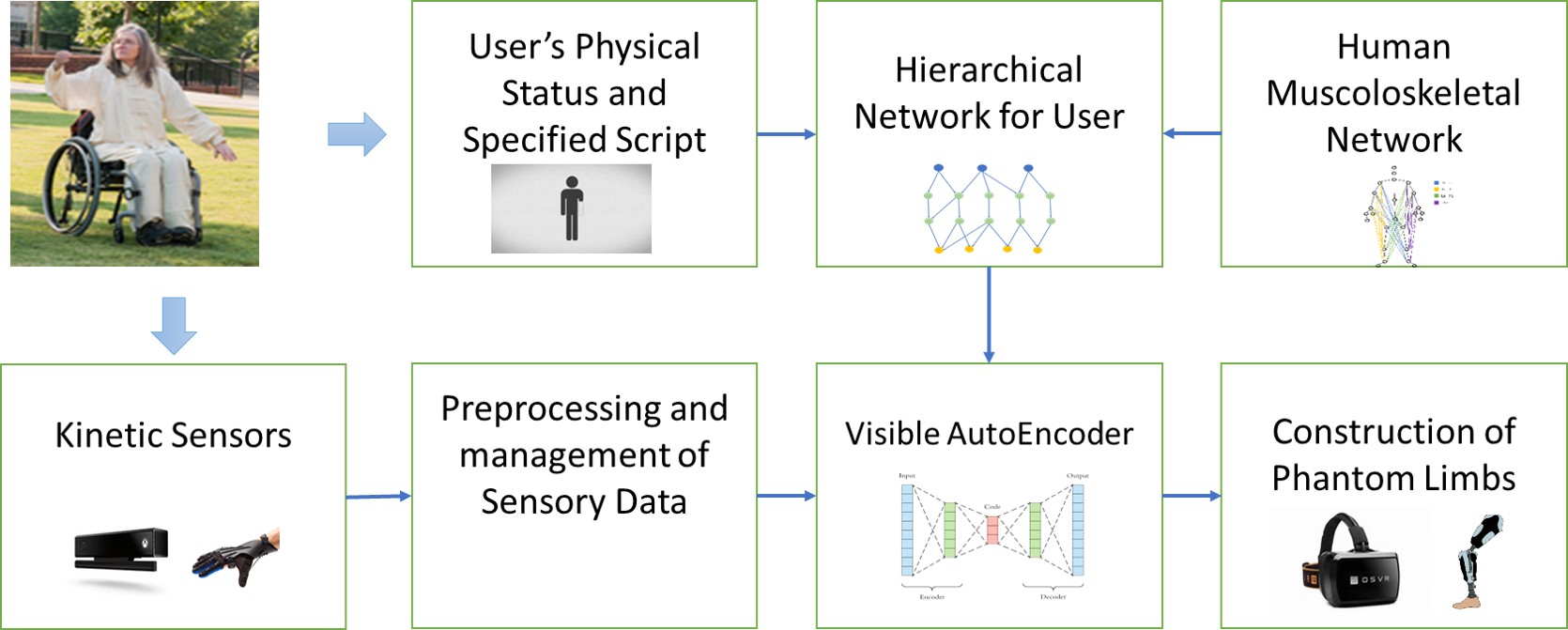}
		\hspace{1cm}
	\caption{The flowchart of the proposed adaptive phantom limb generation based on multi-correlation hierarchical autoencoder}
	\label{Fig2-flowchart}
\end{figure} 

Figure \ref{Fig2-flowchart} shows the implementation of the proposed work, which consists of the following critical tasks:
\begin{itemize}
    \item{Formulating human musculoskeletal network (illustrated in Figure \ref{Fig01-derivation-visible-autoencoder}(a)) \cite{Murphy2018-MusculoskeletanNetwork} according the functional, mechanical and nervous correlation between each body components (muscle, joint, or bone).}
    \item {Deriving hierarchical network (illustrated in Figure \ref{Fig01-derivation-visible-autoencoder}(b), in the configuration of forest data structure) from the human musculoskeletal network according to the physical status of users, where the phantom limbs will form the leaves of hierarchical tree.}
    \item {Building visible autoencoder neural network (illustrated in Figure \ref{Fig01-derivation-visible-autoencoder}(c)) according to the hierarchical network so that the kinetic behavior can be constructed according to the kinetic behavior of user's functional body parts measured by heterogeneous sensors.}
    \item{Preprocess the kinetic sensory data using noise removing, data normalization, incomplete data compensation such as Kalman filter and Spherical linear interpolation (SLERP)\cite{Liang2018-Elsevier-SH}, and kinetic signal decomposition such as singular value decomposition (SVD), and wavelet analysis, etc.
    }
    \item{Training the addressed visible autoencoder neural network according to specific human motions script such as walking, jogging, dancing, or any other physical activities.}
    \item{Representing kinematic behavior about phantom limbs using VR/AR, and actuators,  which can directly stimulate users.}
\end{itemize}

\section{Developing Data Acquisition and Transmission Schemes}
\label{Section:AcquisitionTransmissionSensoryData}

 Proper data acquisition schemes will be developed to collect 4D kinetic data with suitable format  from various hardware instruments such as a Microsoft Kinect V2, foot-pressure sensors, actuators, and VR goggles. Realtime application of two-way communications and edge computing are considered to facilitate the human computer interaction in the PLG-HAE sytem.

\begin{wrapfigure}{L}{0.45\textwidth}
  	\centering
\includegraphics[width=0.45\textwidth, height=0.2\textheight]{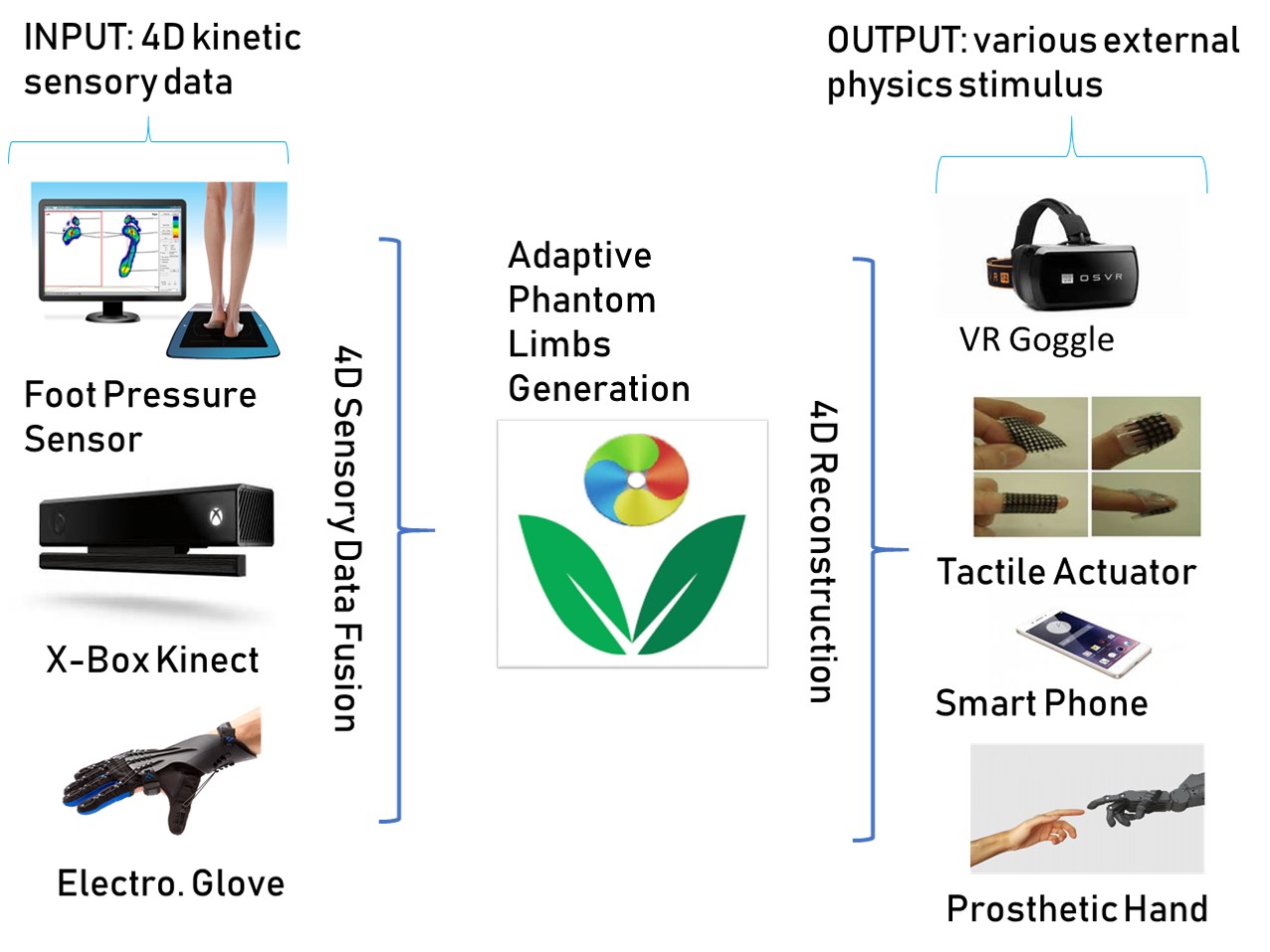}
 	\caption{Input\&output instruments}\label{Fig5-InputOutput}
\end{wrapfigure} 

\subsection{Acquisition of 4D Sensory Data}

Figure \ref{Fig5-InputOutput} show the basic input and output equipment of phantom-limb generation. A Microsoft Kinect V@ sensor, foot pressure sensor, and electronic glove (studded with force-sensitive resistors and vibration motors that can essentially re-create the experience of kinetic movement) can be used to acquire kinetic data (or 4D sensory data) of a user. A virtual reality goggle, such as the Oculus Rift, HTC Vive, or Google Cardboard, motor to drive prosthetic limb, and a tactile actuator are used as output equipment that work together to depict 4D feedback to the user.





\paragraph{ Acquisition and Polishment of Kinematic Data} 
The Microsoft Kinect collects the kinematic data of the coach (or physical therapist, for training purposes) and the user. 
Through Kinect, we can obtain joints' transient position ${\langle x,y,z \rangle}_t^k$ and corresponding Quaternion rotation \cite{Farebrother2003} ${ \langle \cos(\frac{\theta}{2}), \sin(\frac{\theta}{2}) \vec{v} \rangle}_t^k$, where $\theta$ is an angle around  unit axis $\vec{v}$, $t$ is the time, and $k$ is joint's identifier. Quaternions \cite{Farebrother2003} are considered to represent the rotation of a rigid body in 3D space using four degree-of-freedoms (DOFs).


Quaternions are superior to many other traditional rotation formulation methods because they completely avoid gimbal-lock \cite{Grood1983}. In the addressed phantom-limb generation system, Quaternions will be used in 4D reconstruction over Unity3D platform and acquisition of kinetic signal. On the other hand, as a Quaternion is specified with reference to an  arbitrary axis vector it is not a good choice in rotation recognition. In this work, Euler angles $\langle \alpha, \beta, \gamma \rangle$, which represent the angles rotating around axis Z, X, Y respectively (in some literature it is denoted as $\langle yaw, pitch, roll \rangle$) will be adopted in gesture recognition.




The addressed phantom-limb generation system stores the captured kinetic data in JavaScript Object Notation (JSON) format, which includes joint position (${\langle x,y,z \rangle}_t^k$), quaternion rotation (${ \langle \cos(\frac{\theta}{2}), \sin(\frac{\theta}{2}) \vec{v} \rangle}_t^k$), tracking status (0: invisible; 1: referred; 2: observable). Potentially, force load ($\mathbf{f}_t^k$) and momentum, etc. may be included. Tracking status indicates whether or not the specific joint is observable by the depth sensor. The force load and momentum are derived by inverse dynamics analysis. Besides JSON, a kinetic dataset in Comma-Separated Values (CSV) format is also provided in PLG-HAE as many data analytics systems do not accept JSON format.



Due to measurement error or unavoidable occlusion,  a joint is not always observable or tractable by the kinetic sensor. Spherical linear interpolation (commonly abbreviated as SLERP) \cite{Shoemake1985} and  Kalman filtering techniques (be discussed in Section \ref{Section:Preprocessing}) will be employed to compensate the missing data. As illustrated in our preliminary online video \cite{online8}, SLERP can effectively address those short-term missed-tracking joints (namely tracking status=0 or 1). 


\paragraph{Acquisition of Tactile Data}
Besides the Kinect, other acquisition instruments such as an accelerometer, orientation sensors, and strain gauges \cite{Tiwana2012} will also be considered for the addressed system. As indicated above, a foot pressure sensor will be used to obtain the ground reaction force $F_t$  for inverse dynamic analysis. Furthermore, electromyography  (EMG) \cite{Tiwana2012} will be selectively employed to evaluate and record the electrical activity produced by skeletal muscles.
%
%
The EMG signal is characterized by a frequency range of several hertz to over 1 kHz and by amplitudes ranging from fractions of a microvolt to a few thousand microvolts. Electromyogram can be analyzed to detect activation level or to analyze the biomechanics of users' movement. 
%
For the acquisition of high-quality EMG signals from localized muscle region, the PIs will focus on identification of localized muscle region of users, noise reduction and grounding practices (to eliminate extraneous electrical noise), electrode site preparation and placement (to minimize the detection of irrelevant bioelectrical signals) and appropriate differential signal preamplification and preliminary signal conditioning (to further enhance signal-to-noise ratio).

 \vspace{-2mm}
\paragraph{Reconstruction of 4D data}
 4D kinetic feedback/instruction is reconstructed through virtual reality and tactile actuators. For example, HTC Vive can be used as the VR device, as it is readily supported and available to consumers and does not exhibit poor VR qualities, such as low refresh rate, that may induce nausea to the user. Unity3D may be used as the SDK for generating a virtual environment using AR and VR. To visualize the human body through a virtual reality facility, the output kinematic data is translated into Quaternion format \cite{Farebrother2003,Grood1983}, which is recognized by Unity3D. The conversion of joint input data to Quaternions allows for the manipulation of a 3D human model, through the manipulation of a model's child objects or joints. The manipulation of a model allows for a better visual representation of a body in both recorded and real-time Kinect data.


Equipped with somatosensory stimuli, the phantom limb generator also directly guides users with somatosensory feedback. Tactile actuators potentially used in this work include Eccentric Rotating Mass (ERM), Linear Resonant Actuator (LRA), Piezo, and Electro-Active polymers (EAP). EAP will be investigated in the development of phantom-limb generation because of its 
high fidelity of sensations, and excellent durability.

\begin{wrapfigure}{L}{0.43\textwidth}
\centering
\includegraphics[width=0.43\textwidth]{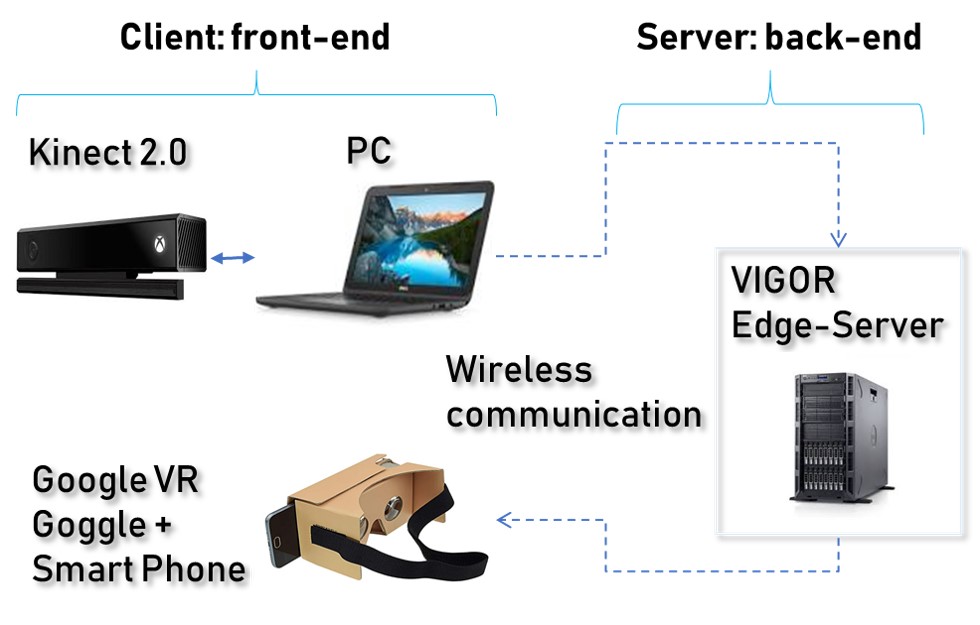}
	\caption{Edge-computing-enabled phantom-limb generation deployed on commodity hardware (demo in online video  \cite{online12-Dakila-EdgeComputing2018})}
	\label{Fig8-edge-computing-GPL-HAE}
\end{wrapfigure}

We will apply deep learning techniques to electric prosthesis with the goal of developing an agent that could learn and adapt to the needs of the user and changes in the environment \cite{Vasan2017-Prosthesis}. Deep learning provides a nice set of tools that allow us to incorporate numerous sensory streams (the kinetic behavior of other body parts, vision, tactile sensation,  inertial measurements, servo position, etc.) in ways that are computationally tractable and suitable for online learning. In this work, autoencoder neural network enables the phantom limbs coordinate with other body parts according to their correlation, some latency is observed. In the future work, more "intelligent" deep learning techniques such as reinforcement learning will be investigated to improve the performance of prosthetic limb (with much less latency).

\subsection{Task: Developing Communication and Edge-Computing Protocols}
\label{section:communication}



\vspace{2mm}
\paragraph{Real-time, Two-Way Communication.}
 Two-way communications are of key importance in the proposed system, since the information needs to be exchanged in a real-time manner. The challenges of the communication protocol for the proposed system include:
(1) Real-time communication: Information in the addressed system needs to be conveyed in real time. If there is a significant delay in the communications, synchronization between the virtual coach and user will be lost and the user will experience a disjointed rhythm. 
(2) High throughput: When there are many users, all corresponding video and audio need to be conveyed in the network, thus incurring a substantial requirements for communication bandwidth.
(3) Two-way communications: The communications are between the virtual coach and users with mutual interactions. Therefore, it could be sub-optimal if one-way communications are considered separately.
(4) Dynamics awareness: The communications may be optimized together with the physical dynamics of the virtual coach and users (namely the motions).

To address the above challenges, we will first model phantom-limb generation as a cyber physical system (CPS) \cite{Lee2008, Sha2008} and then analyze the bandwidth needed for controlling the physical dynamics. Then, the detailed communication protocol will be designed and evaluated with the whole system.

\vspace{-2mm}
\paragraph{Deployment of Phantom-limb Generation Using Affordable Hardware Based  on Edge Computing.}
Edge computing enables real-time knowledge generation and application to occur at the source of the data close to user device \cite{Hu2015, Mkinen2015}, which makes it particularly suitable for the proposed latency-sensitive system. An edge server can be adapted to serve multiple users through interaction with their devices. There are communication and computing trade-offs between the edge server and each user device. Data could either be locally processed at the user device or else be transmitted to and processed at the edge server. Different strategies introduce different communication costs, resulting in a difference in delay performance. To provide the best quality of experience for users, the PIs will conduct the following studies for the proposed PLG-HAE system: (1) Identification and modularization of computing tasks: the computing tasks of data preprocessing, kinetic movement recognition, the analysis of individualized movement choreography and the corresponding computing overheads (CPU cycles, memory) will be determined. (2) Design, prototyping and enhancement of offloading schemes: Based on the results of bandwidth and delay analysis as well as delay performance requirement, computation offloading schemes will be developed to determine which computing tasks should be performed locally at the user device and which computation tasks should be offloaded to the edge server.


As illustrated in Figure \ref{Fig8-edge-computing-GPL-HAE} (b), enabled by edge-computing technology, the proposed phantom limb generation system can be deployed over commodity hardware so that it can benefit a broader population.
An illustrative concept demonstration about edge-computing-enabled phantom limb generator is given in our online video  \cite{online12-Dakila-EdgeComputing2018}.
\section{Deriving Hierarchical Neural Network from Human Musculoskeletal Network}

\subsection{Structure, function, and control of the human musculoskeletal network}

\begin{wrapfigure}{l}{0.3\textwidth}
 	\centering
	\includegraphics[height=5.5 cm]{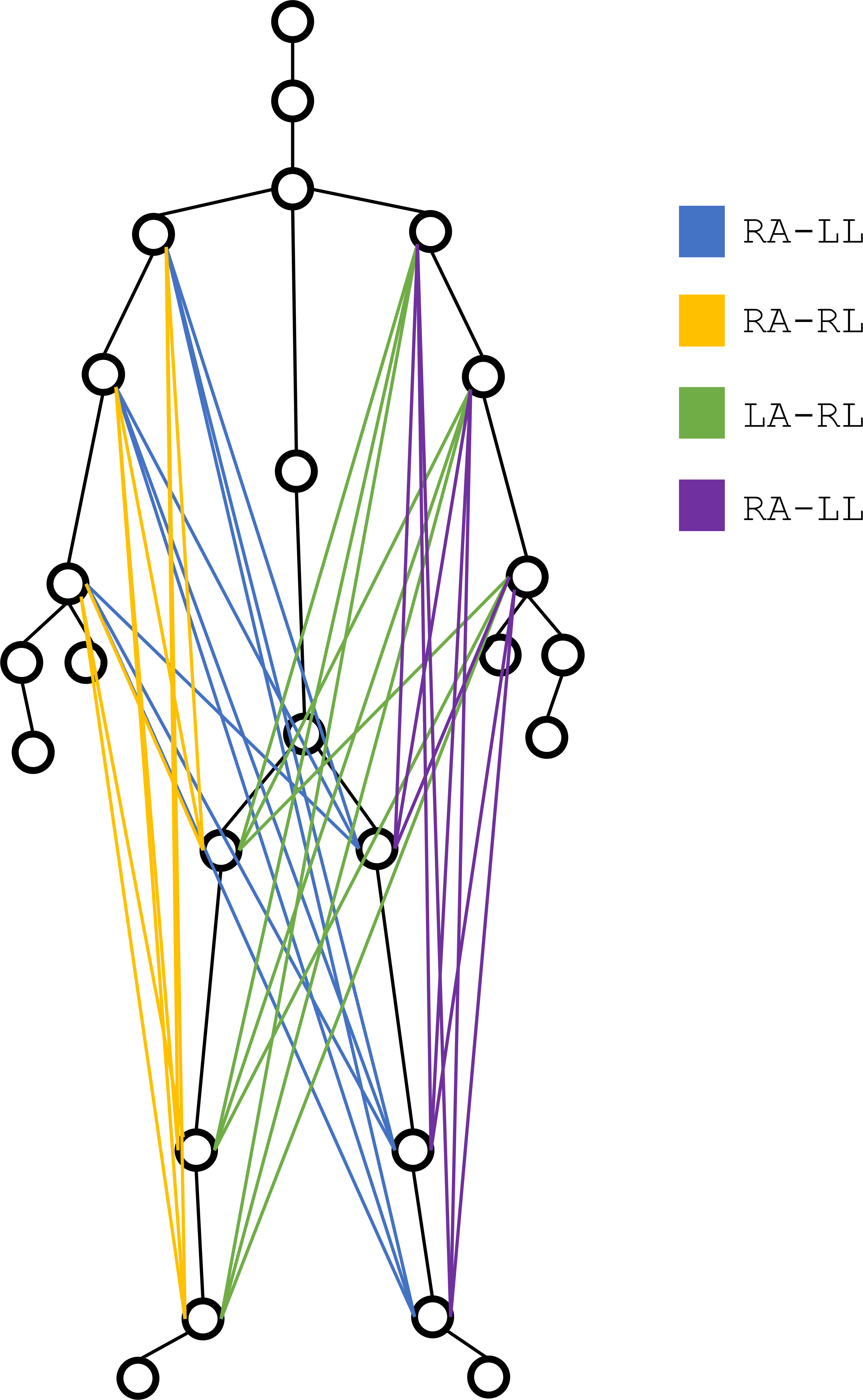}
	\caption{Human musculoskeletal network.}
	\label{Fig06-HumanMusculoSkeletalNetwork}
\end{wrapfigure} 

As illustrated in Figure \ref{Fig06-HumanMusculoSkeletalNetwork}, the human body is a complex organism, the gross mechanical properties of which are enabled by an interconnected musculoskeletal network \cite{Murphy2018-MusculoskeletanNetwork} controlled by the nervous system. The nature of musculoskeletal interconnection facilitates stability, voluntary movement, and robustness to injury. However, a fundamental understanding of this network and its control by neural systems has remained elusive. Here we address this gap in knowledge by utilizing medical databases and mathematical modeling to reveal the organizational structure, predicted function, and neural control of the musculoskeletal system. We constructed a highly simplified whole-body musculoskeletal network in which single muscles connect to multiple bones via both origin and insertion points. We demonstrated that, using this simplified model, a muscle's role in this network could offer a theoretical prediction of the susceptibility of surrounding components to secondary injury. Finally, we illustrated that sets of muscles cluster into network communities that mimic the organization of control modules in primary motor cortex. This novel formalism for describing interactions between the muscular and skeletal systems serves as a foundation to develop and test therapeutic responses to injury, inspiring future advances in clinical treatments.


\subsection{Derivation of hierarchy out of musculoskeletal network}

Human-body consists of multiple body-components, which are correlated functionally, bio-mechanically, or nervously. From the correlation graph, we can derive a hierarchical sub-graph that can be formulate by neural network in a relatively straightforward way.
In this work, hierarchical clustering algorithm \cite{Langfelder2008-HierarchicalClustering} is employed to formulate the hierarchical network for phantom limbs out of human muscoloskeletal network.


\subsection{Visible and hierarchical neural network architecture for real-time phantom limb generation.}

\begin{wrapfigure}{L}{0.6\textwidth}
     \centering
     \includegraphics[width=0.6\textwidth]{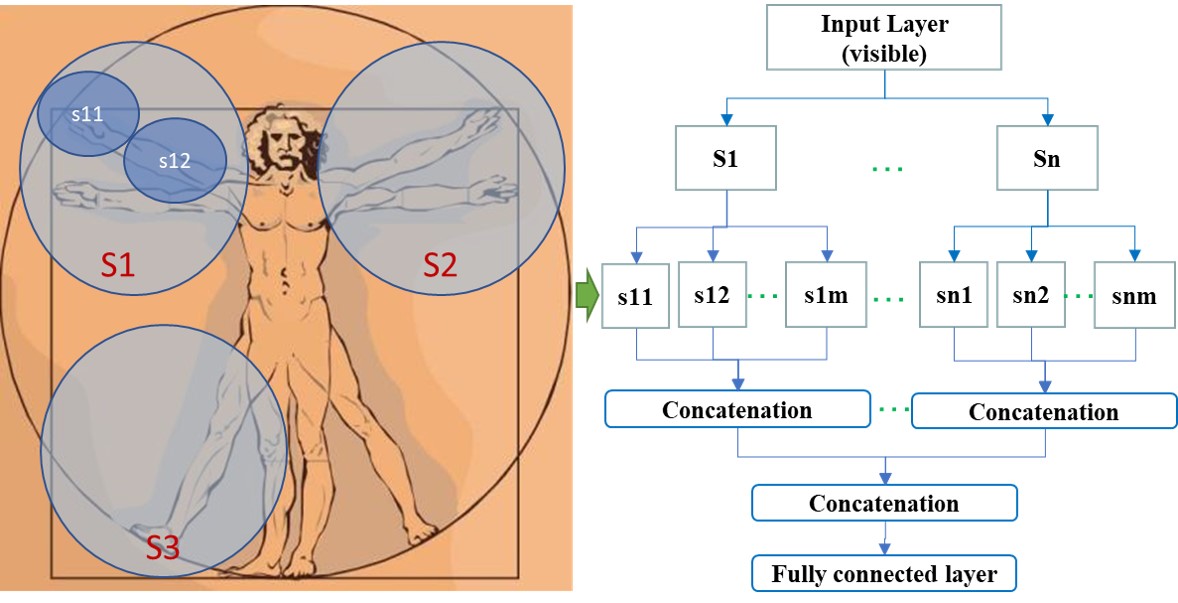}
 	\caption{Visible and hierarchical neural network}
 	\label{Fig3B-VisibleHierarchicalNN}
\end{wrapfigure}
It is known that any system can be regarded as a hierarchical structure (i.e., system $\rightarrow$ subsystem $\rightarrow$ subsubsystem, \ldots). As illustrated in Figure \ref{Fig3B-VisibleHierarchicalNN}(a), the human body system can be always divided into sub-components that are mechanically correlated and whose motions are non-divergent. Inspired by the Bayesian network, we propose a visible and hierarchical neural network to accurately formulate a system. As illustrated in 
Figure \ref{Fig3B-VisibleHierarchicalNN}(b), a sample visible and hierarchical neural network, which is directly derived from the human body system, is employed to specify the musculoskeletal kinematic. The visible and hierarchical neural network can be employed in phantom limb generation, 4D kinetic behavior recognition, and individualized Tai-Chi choreography (to be discussed in the remaining sections). 
Preliminary experimental results demonstrate that is superior to a classical neural network from the point of view of training speed and stability. 

\subsection{Related Work}
Much of work done in human motion generation focuses on choosing a model expressing the inter-dependence between joints. Recently, human motion was generated through the use of auto-regressive functions, such as models relying on Gated Recurrent Networks (GRU) or Long-Short Term Memory (LSTM) \cite{Goodfellow2016}. However, these models are not feasible for use in our applications. Unlike long-term or short-term prediction, models focusing on real-time generation must be small and computationally simple to achieve a 24 predictions-per-second minimum. Other work has focused on minimizing loss by using techniques such as highway units, custom loss functions, or defining loss parameters. Generative-Adversarial Networks (GAN) \cite{Goodfellow2016} have also been experimented. However, these models tend to be complex and convergence of the neural network is not intuitive. Thus, determining ground truth error for these models is nuanced during training. Due to the inability to use these neural networks, methods including real-time autoencoders are examined in this paper.

\begin{figure}[htp]
\centering 
	\includegraphics[width=0.9\textwidth]{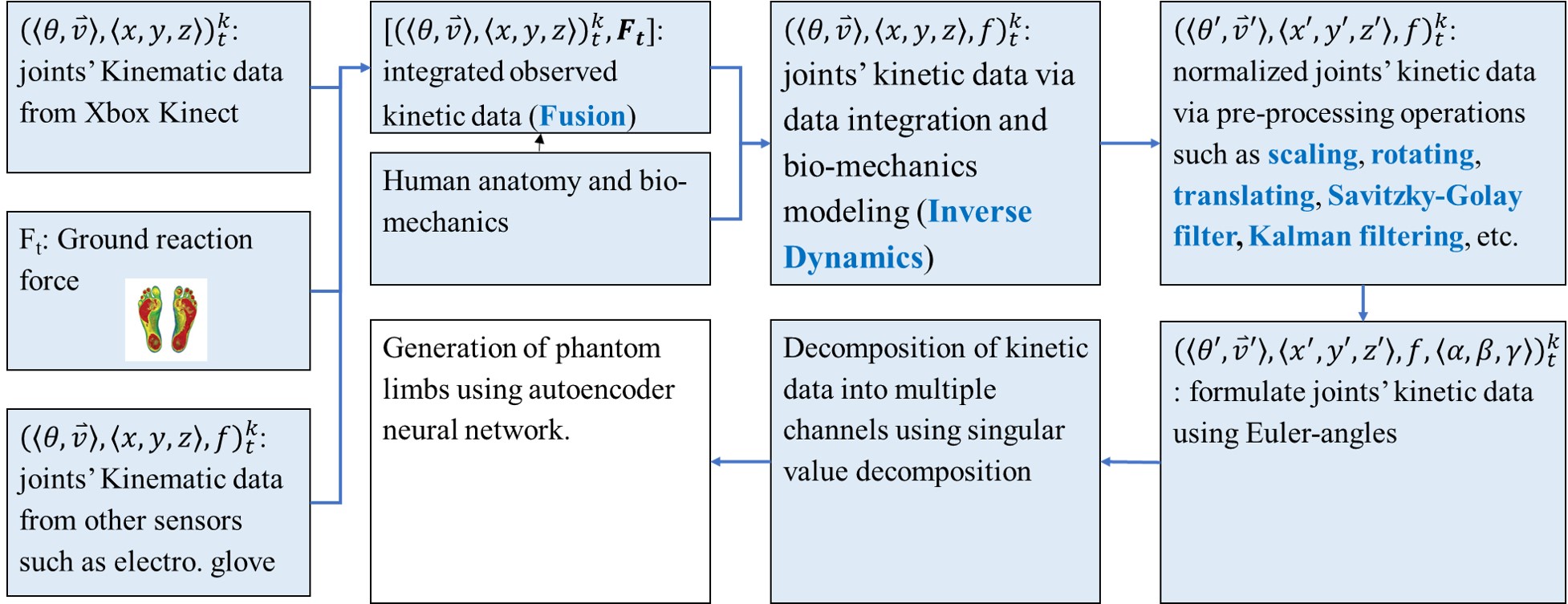}
	\caption{Flowchart of the Preprocessing for the generation of phantom limb generation}
	\label{Fig9-PreprocessingFlowchart}
		\vspace{-4mm}
\end{figure}

\section{Developing Data Preprocessing Methods of Autoencoder-enabled Phantom Limb Generation}
\label{Section:Preprocessing}

Data preprocessing operations play an indispensable role in phantom limb generation because: 
     (1) {\em Input data is of a heterogeneous nature}.
     For example, different users have variable sizes; sensors may have various viewing angles; users will not always be located in a deterministic position; and the two time-series data sets may not be synchronized. As a result, scaling, rotating, translating, and dynamic time warping (DTW) are needed to normalize the original input data.
     (2) {\em The input data set may be incomplete}. For example, occlusion inevitably leads to missing data; Musculoskeletal force and momentum exerted over the joints or muscle cannot be directly obtained from the sensors; some input channels are not enabled (e.g., partial control) for users with mobility-based chronic conditions (i.e., partial control). In the implementation of autoencoder-enabled phantom limb generation, Kalman filtering, inverse dynamics, and time-series prediction are employed to handle the incomplete data \cite{Alharbi2017,D.Ledesma2018}.
     (3) {\em The measurement-induced noise is significant}.


Figure \ref{Fig9-PreprocessingFlowchart} shows the flowchart of data preprocessing of autoencoder-enabled phantom limb generation \cite{Liang2015,C.Davis2018,online1,online2}. $\langle \theta, \vec{v} \rangle$ (denoted as $ \langle \cos{\frac{\theta}{2}}, \sin{\frac{\theta}{2}} \vec{v} \rangle $ in Section \ref{Section:AcquisitionTransmissionSensoryData}, where $ \theta $ is the rotation angle about axis $\vec{v}$ ) indicates a joint's Quaternion rotation; $\langle x,y,z \rangle$ denotes a joint's position; $f_t$ indicates a joint's applied force, which is derived from inverse dynamics; $\langle \alpha, \beta, \gamma \rangle$ indicates a joint's rotation under normalized Joint Coordinate System (JCS) --  Euler angle.   Its main implementation techniques include data fusion, inverse dynamics analysis, spatial normalization, Kalman filtering, and reconstruction of disable input channels. The kinetic data is stored in JSON format.



\subsection{Formulating Musculoskeletal Kinetic Features }

Inverse dynamics analysis (IDA), which is derived from Newton-Euler equations \cite{Ambrosio2007,Eich-Soellner2008,Moreira2013}), aims to calculate unknown kinetic information (the force and moments of joints and muscles) from measured kinematic information (e.g., position, velocities and accelerations of joints) and measured kinetic information (e.g., ground reaction force measured by foot pressure sensor). As illustrated in Figure \ref{Fig9-PreprocessingFlowchart}, given joints' location $\langle x_i,y_i,z_i \rangle $, Euler rotation $\langle \alpha_i,\beta_i,\gamma_i \rangle$ where $i$ denotes the identity of a joint, and ground-reaction force $F$, the joints' force $f_i$ and other musculoskeletal kinetic features can be computed via IDA.


As illustrated in Figure \ref{Fig9-PreprocessingFlowchart}, autoencoder-enabled phantom limb generation employs inverse dynamics to compute internal joint forces and moments with given ground reaction forces. 
In this work, the human body is divided into multiple connected rigid bodies \cite{Liang2013-FEM, Wittenburg2008} which correspond to relevant anatomical segments such as the thigh, calf, foot, arm, etc. The model's anthropometric dimension (e.g., the mass and momentum inertia) is derived from statistical analysis. In addition, it is assumed that each joint is rotationally frictionless.

The proposed methods  in Figure \ref{Fig9-PreprocessingFlowchart} can be customized to investigate the bio-mechanical response of human motion by considering appropriate pathology for different health issues such as cerebral palsy, poliomyelitis, spinal cord injury, and muscular dystrophy \cite{Moreira2013}. This will make the addressed work a more constructive rehabilitation system.

\subsection{Spatial Normalization}

As addressed in Section \ref{Section:AcquisitionTransmissionSensoryData}, we can acquire the joints' position and rotation, which are denoted as $\langle x, y, z\rangle$ and $ \langle \theta, \vec{v} \rangle $ respectively. Both need to be normalized to ease and boost the gesture recognition:

\paragraph{Normalization of Joints' Rotation}
Spatial rotations in three dimensions can be parametrized using either Euler angles and unit quaternions. Instead of being denoted as a rotation around an arbitrary vector such as a unit Quaternion, an Euler angle describes the orientation of a rigid body with respect to a fixed coordinate system. Therefore, in kinetic gesture recognition, Euler angle is more constructive than Quaternion in data analytics.
In the proposed research, joints' rotations can be normalized by converting the quaternion $\langle \theta, \vec{v} \rangle$ into Euler angle $\langle \alpha, \beta, \gamma \rangle$.

\paragraph{Normalization of Joint's Position}
Joints' position $\langle x, y, z\rangle$ involves many factors irrelevant to  gesture recognition. These include: a user's orientation to the camera, location in the room, and body shape and size. To eliminate such irrelevant factors, 
This work employs a series of spatial normalization techniques: (1) bone scaling, which makes uniform the bone length of users \cite{Alharbi2017}; (2) axis-oriented rotating of view angle; (3)  translation of the origin, which makes a user  positioned at the center of a sensor; (4) re-constructing $\langle x, y, z\rangle$ according to joint rotation \cite{Grood1983} 
; and (5) polishing the kinetic curve using a Savitzky-Golay filter.
Our preliminary experimental results demonstrate that the normalization techniques addressed above can greatly improve the quality of data (less noise and smoother kinetic performance) so as to achieve higher recognition \cite{Alharbi2017,C.Davis2018,Liang2015}.

\subsection{Recovering Occlusion-induced Missing Data}


During  sensory data acquisition, unavoidable occlusion may introduce missing data or lost-tracking. As addressed in Section \ref{Section:AcquisitionTransmissionSensoryData}, SLERP can basically fix the issues caused by short-term occlusion.  Kalman filter \cite{Kalman1960, Einicke2012, Zarchan2000} can be employed 
to fix the missing information (including both position and rotation) caused by long-term occlusion. 
A preliminary comparison between the raw and preprocessed physical rehabilitation kinematic data is available on our online video \cite{online1, online2}. 


\subsection{Application of Singular Value Decomposition on Kinetic Data}

Singular value decomposition (SVD) is employed to decompose the human kinetics into multiple of modes, which will be fed into the autoendoer in different channels. SVD can reduce the nonlinearity degree of kinetic signal so that it can be formulated by neural network more precisely.
\section{Neural Network Architecture and Generation of Phantom Limbs}
\subsection{Previous Work}
Deep neural networks (DNN) have proven to be significant in feature learning through their ability to learn representations in a hierarchical manner \cite{butepage2017-deepmotionnn}. Specific to the applications focusing on the generation of human motion data, DNN models based on recurrent neural networks (RNN), such as ones that employ Long-Short Term Memory (LSTM) and Gated Recurrent Units (GRU) currently represent the majority of the techniques proposed. These proposed models, such as ones that focus on temporal encoding, were more focused on learning general, generic representations from a large pool of general, non-specific human motion data. Thus, previous models were more focused on the generation of new, novel motions. 

In contrast, our work is focused on assessing current body language that results in the representation of missing joints given a deterministic motion. In addition, unlike previous models that rely on large datasets, we must be able to train on very small datasets (such as ones with less that 25 samples per motion). Many architectural modifications are proposed to attain proper performance given these constraints.

For the recreation of missing data through phantom limbs, two traditional methods for machine learning were assessed -- multilayer perceptron (MLP) and denoising autoencoders (DAE). In addition, this we proposed a new autoencoder architecture for the recreation of phantom joints, a multi-correlated hierchical autoencoder. 

\subsection{Generating phantom legs based on arm movement using visible and hierarchical autoencoder network.}

As our preliminary contribution, a neural network is trained to generate the kinetic status of hip, knees, and feet according to the kinetic status of shoulders, elbows, and arms captured by 4D sensors \cite{online13-Dakila-Autoencoder-PhantomLeg2018}.
As illustrated in figures \ref{daxfigure-mlp-arch}, \ref{daxfigure-sdae-arch}, \ref{daxfigure-dae-split-arch}, and \ref{daxfigure-h-autoencoder}, four network architectures are investigated in this research: (a) multiple layer perceptron (MLP); (b) denoising autoencoder (a classical autoencoder architecture); (c) visible and hierarchical neural network with two subsystems (VHNN2); and (c) VHNN with four subsystems (VHNN4). 

It can be observed that VHNN splits the input tensor and then feeds the split tensor into multiple smaller, parallelized autoencoders. Thus, data for each joint can be calculated in parallel with their own respective autoencoder. The aforementioned parallelized autoencoder pipelines are simplified stacked autoencoders, allowing for optimization of specific, key tasks rather than one large task.

A video playlist of the generation of phantom legs based on VHNN may be found at 
\cite{online20-Dakila-PhantomLeg2018}.

\subsection{Multilayer perceptron models}

Multilayer perceptrons (MLP) were assessed as a preliminary, prototype network for the inference and generation of phantom joints. Despite their simplicity, MLP models are able to learn non-linearly separable data, and can learn features through stochastic approximation. Thus, they are the simplest deep model available, and was used as a minimum benchmark to the other models we have thus developed. Rather than feeding input into a logistic regression, MLP networks work by adding layers that process the input using activation functions. 

\begin{figure}[htp]
 	\centering
 	\includegraphics[width=15.5 cm]{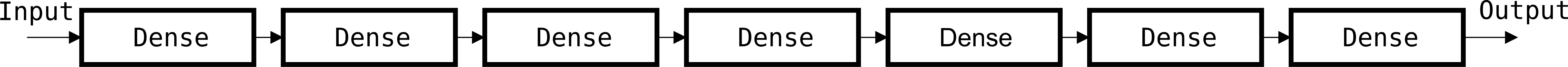}
	\caption{Generation of phantom legs from moving arms using a multilayer perceptron architecture \cite{online20-Dakila-PhantomLeg2018})
	}
	\label{daxfigure-mlp-arch}
\end{figure} 

As seen in Figure \ref{daxfigure-mlp-arch}, our implementation of a MLP Regressor model was built with 7 hidden layers, with each layer possessing 18 nodes. The MLP model is also fitted with a regularizer term to prevent over-fitting of our motion data. However, the complexity of our data proved too difficult for the model to learn, and thus our implementation of an MLP regression model achieved poor accuracy \cite{online14-Dakila-MLP-PhantomLeg2018} as the model struggles to generate proper regression values. The increase and/or change of hyperparameter values, such as layer number or nodes per layer, do not significantly benefit accuracy of the results, but do increase the training time.

\subsection{Denoising Autoencoder}
In generative data applications, autoencoders have had an important influence for their low computation cost--relative to RNNs--and their high accuracy. For the generation of phantom limbs, autoencoders (AE) were critically assessed in their performance for generating human-like motion data whilst being quick in their predictions/generation of data. Quickness and speed of prediction is important to the generation of phantom limbs as the predictions must be done in realtime, i.e. generated over 24 times per second. Whilst the generation of temporal/sequence data are most normally done through recurrent neural networks, convolutional neural networks such as autoencoders have found success in generative tasks \cite{qiao2019-hierchicalcnnwords}. During initial testing, autoencoders exhibited data bias within the generated movement, causing difficulties in the neural network converging \cite{online20-Dakila-PhantomLeg2018}. 

For phantom limb generation, autoencoders are quite applicable as the generation of data does not necessarily have to be over a large time window. Instead, accuracy during short-time windows and quick reaction speed relative to the realtime input were more important. In addition, convolutional neural networks are less susceptible to noisy or substandard data and can better generalize data compared to the more complex recurrent units.

\begin{figure}[htp]
 	\centering
 	\includegraphics[width=15.5 cm]{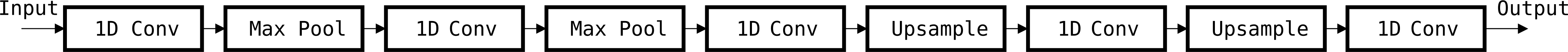}
	\caption{Generation of phantom legs from moving arms using a classical denoising autoencoder architecture \cite{online20-Dakila-PhantomLeg2018})
	}
	\label{daxfigure-sdae-arch}
\end{figure} 

As seen in \ref{daxfigure-dae-split-arch}, we moved our attention away from MLP and employed a standard denoising autoencoder to learn the correlations of one's body language to a specific set of phantom limbs. Due to the noisy nature of data captured by our Microsoft Kinect V2, the performance of autoencoders for denoising as well as generative representation was assessed. Through the use of a stacked denoising autoencoder, accuracy when viewed in comparison to multi-layer perceptron models were significantly higher in terms of the motion data generated and exhibited less noise within the movement \cite{online14-Dakila-MLP-PhantomLeg2018}\cite{online20-Dakila-PhantomLeg2018}. In contrast to data generated by MLP models, autoencoders achieved motion data that was believably human-like.

\subsubsection{Autoencoder Data Bias}
Despite the improvement of generation of human-like data using autoencoders, the generated motions were not perfect for all phantom limbs. Using our implementation of a denoising autoencoder, the minimization of loss presented bias towards data points that were closer to the starting joints. Our autoencoder's input and output tensors exhibit a dimensionality of (n, 18, 3), representing sample number, joint index, and x, y, z-coordinates respectively. The generation of motion data found within the first few joints, such as the first six indices out of 18, exhibited much better human motion data in comparison to data found in the last twelve data points. Thus, motion generation data generated for a specific leg (such as the left leg) was significantly more believable than data generated for the complement leg (such as the right leg). The exhibition of the bias in optimization is most apparent during the realtime generation/prediction of data \cite{online20-Dakila-PhantomLeg2018}.

The exhibition of data bias may due to the hierchical nature of the human body, e.g. the musculoskeletal network. Due to the divergent motion relation of different body segments, optimization of the may be compromised when other joints exhibit opposite or very divergent motion. Thus, the correlations between one's body language may be intuitive for joint segment (i.e. one's left leg), but entirely opposite or non-intuitive for another joint segment (i.e. one's right leg).

\subsection{Application of Hierarchical Autoencoder}
The prior applications of hierchical autoencoders can be found in classification tasks and natural language processing (NLP) \cite{Li2015-hierchicalaewords}. In retrospect, these neural networks have been employed to decrease complexity within ambiguous classifications \cite{Yian2019-hierchicalcnnfashion} or used as a means to assess long sequences of data using CNNs \cite{qiao2019-hierchicalcnnwords}. The benefits of using hierarchical CNNs for sequential data are two-fold: (1) They are much quicker to train and predict and (2) They can attain higher accuracy than RNNs in certain applications. 

We turned to the parallelization of convolution pipelines for the generation of phantom limbs in response to the optimization problems by our stacked denoising autoencoder showed. Thus, the attainment of certain goals was needed for the new, parallelized autoencoder models: (1) the autoencoder must not too complex that the generation of data would not be realtime, (2) the autoencoder must not exhibit signs of optimization bias, and (3) the autoencoder must be able to allow for multi-correlation of outputs.

\begin{figure}[htp]
 	\centering
 	\includegraphics[width=15.5 cm]{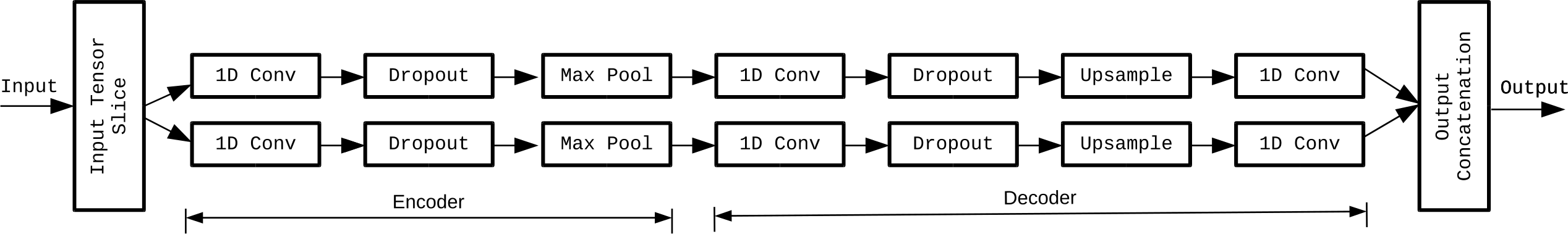}
	\caption{Generation of phantom legs from moving arms using a two thread (subsystem) visible and hierarchical autoencoder neural network (VHNN-AE-2) architecture \cite{online20-Dakila-PhantomLeg2018})
	}
	\label{daxfigure-dae-split-arch}
\end{figure} 

Figure \ref{daxfigure-dae-split-arch} shows the autoencoder architecture used to resolve this inherent data bias. As seen, the proposed network has its splits each input tensor according to the joint joint segments that the data is correspondent to for both input and output, in comparison to figure \ref{daxfigure-sdae-arch}. This resolves biased optimization by allowing each autoencoder to optimize for a smaller input data dimensionality, and for motion that is mostly non-divergent. Because of the reduced dimensionality of input data, the MaxPooling as well as Upsampling of data does not need to encode and decode as large of data in comparison to a deeper, larger stacked autoencoder. Thus, despite the seemingly double amount of layers, the amount of layers per smaller autoencoder is reduced, and parameters per parallelized autoencoders are also reduced. This results in training and prediction performance comparable to a standard denoising autoencoder, but with much higher ground-truth accuracy.

\begin{figure}[htp]
 	\centering
 	\includegraphics[width=15.5 cm]{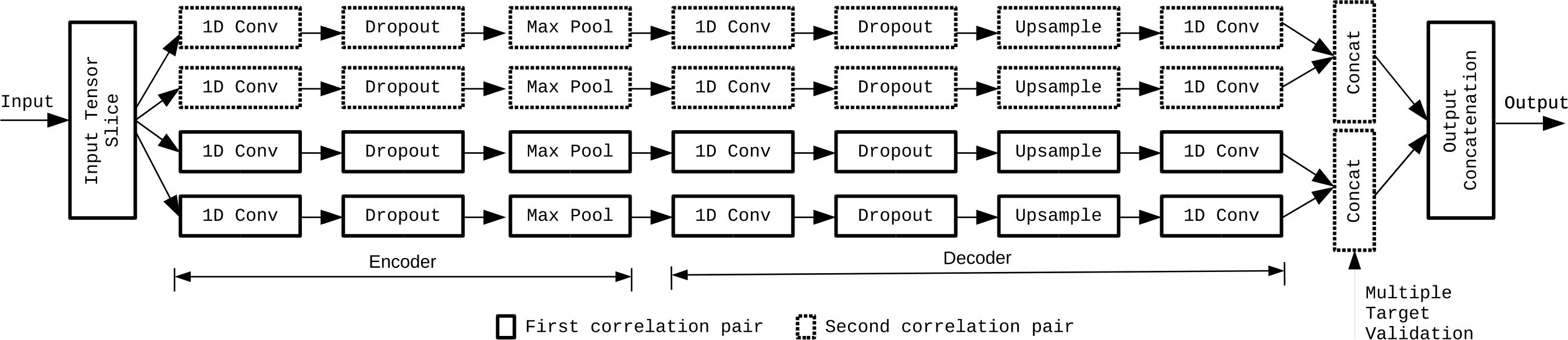}
	\caption{Generation of phantom legs from moving arms using a four thread (subsystem) visible and hierarchical autoencoder neural network (VHNN-AE-2) architecture \cite{online20-Dakila-PhantomLeg2018})
	}
	\label{daxfigure-h-autoencoder}
\end{figure} 

Whilst the splitting of inputs into two parallel autoencoders performed much better than a standard stacked autoencoder, the parallelization of autoencoders has caused a separate problem, in which correlations can only be found in two distinct joint pairs, i.e. one's left arm motion can influence the left leg but not the right leg. Thus, multi-correlation was needed in order to help give better representation of each input to the generated output. As seen in Figure \ref{daxfigure-h-autoencoder}, this resulted in further splitting of the autoencoders, essentially doubling the amount of pipelines within this phantom limb application. This was done to give correlations to both input joints segments to both output joint segements. 


\begin{table}[htp]
\centering
\caption{Architectures of Autoencoders Assessed and Proposed}
	\begin{tabular}{|l|l|l|l|l|l|} \hline
	\multicolumn{2}{|c|}{Classical} & \multicolumn{2}{|c|}{VHNN-2}   & 	\multicolumn{2}{|c|}{VHNN-4} \\ \hline
    Type   & Hyperparameters   & Type       & Hyperparameters    & Type       & Hyperparameters \\ \hline
    conv1D & Filter: 32        & conv1D     & Filter: 32         & conv1D     & Filter: 32     \\ 
           & Kernel Size: 3    &            & Kernel Size: 3     &            & Kernel Size: 3 \\
           & Activation: Linear&            & Activation: Linear &            & Activation: Linear \\
           & Padding: same     &            & Padding: same      &            & Padding: same \\ \cline{3-6}
           &                   &  Dropout   & Rate: 0.5          & Dropout    & Rate: 0.5     \\ \hline
MaxPooling & Pool size: 3      & MaxPooling & Pool size: 3       & MaxPooling & Pool size: 3     \\ \hline
    conv1D & Filter: 32        &            &                    &            &                  \\ 
           & Kernel Size: 3    &            &                    &            &                  \\
           & Activation: Linear&            &                    &            &                  \\
           & Padding: same     &            &                    &            &                  \\ \cline{1-2}
MaxPooling & Pool size: 2      &            &                    &            &                  \\ \cline{1-2}
    conv1D & Filter: 32        & conv1D     & Filter: 32         & conv1D     & Filter: 32     \\ 
           & Kernel Size: 3    &            & Kernel Size: 3     &            & Kernel Size: 3 \\
           & Activation: Linear&            & Activation: Linear &            & Activation: Linear \\
           & Padding: same     &            & Padding: same      &            & Padding: same \\ \cline{5-6}
           &                   &            &                    & Dropout    & Rate: 0.2     \\ \hline
 Upsampling& Size: 2           & Upsampling & Size: 3            & Upsampling & Size: 3       \\ \cline{1-2}
conv1D     & Filter: 32        &            &                    &            &            \\ 
           & Kernel Size: 3    &            &                    &            &            \\
           & Activation: Linear&            &                    &            &            \\
           & Padding: same     &            &                    &            &            \\ \cline{1-2}
 Upsampling & Size: 3          &            &                    &            &         \\ \hline          
conv1D   & Filter: 1        & conv1D     & Filter: 1         & conv1D     & Filter: 1     \\ 
           & Kernel Size: 3    &            & Kernel Size: 3     &            & Kernel Size: 3 \\
           & Activation: Linear&            & Activation: Linear &            & Activation: Linear \\
           & Padding: same     &            & Padding: same      &            & Padding: same \\ \cline{3-6}
           &                   &Concatenate & Axis: 1            & Concatenate & Axis: 1               \\ \cline{5-6}
           &                   &            &                    & Concatenate&  Axis: 1         \\ \hline
	\end{tabular}
\label{Table4-VHNN-architecture}
\end{table}

{\em Partial control} is an important feature of the generation of phantom limbs as it enables those users with disability to operate freely. Partial control is implemented by compensating the missing data inferred by disabled input channels: 
in the event that several input channels are disabled,  the addressed model is able to construct the void input channels by taking the advantage of correlation among all inputs. 
Compensation can normalize the input data so that VIGOR can achieve higher recognition rate, its psychological and physiological benefits to users are also under our investigation.

%


\section{Results}
\label{subsection:compensationMissingData}
\vspace{2mm}

\begin{table}[htp]
\centering
\caption{Time-performance of phantom-legs generation using visible and hierarchical autoencoder neural network (VHNN) (Intel Core i9-7900X, 1x NVIDIA GTX1080 Ti, 64GB RAM; MLP does not employ GPU)}
	\begin{tabular}{|l|l|l|l|l|l|}
		\hline
		NN architecture	 & Training time  & Training time & convergence    & ground      & online \\ 
		                 & per step       & for 1K epochs & in epoch      & truth error  &     video   \\ \hline
		MLP					  & NA   			 & NA       &  ~250  &   9515.51  &  \cite{online14-Dakila-MLP-PhantomLeg2018}\\
		Denoising Autoencoder & $108-110\mu s$   & 9m 15s   &  ~1000 &   2107.46  & \cite{online15-Dakila-MLP-PhantomLeg2018}\\
		VHNN-Autoencoder-2	  & $30-31\mu s$	 &  2m 44s      &  ~500     &   276.68       & \cite{online16-Dakila-MLP-PhantomLeg2018}\\
		VHNN-Autoencoder-4    & $52-57 \mu s$	 & 4m 37s   &  ~800  &   366.15   & \cite{online17-Dakila-MLP-PhantomLeg2018}\\ \hline
	\end{tabular}
\label{Table5-VHNN-phantom-legs-generation}
\end{table}

As illustrated in Table \ref{Table5-VHNN-phantom-legs-generation}, the proposed VHNN 
 architecture has proven to have overall superior results compared to previous work. \textcolor{black}{Decreased training time compared to previous autoencoders architectures can be observed due to the parallelization of simpler autoencoders, increasing efficiency by easing optimization. This is done by allowing autoencoders to train on specific gestures in a whole movement. In addition, it does not exhibit data-hungry tendencies that state-of-the-art models exhibit, allowing it to be trained on small amounts of data.}
 
 \textcolor{black}{Lower ground truth error can be seen in the VPNN-AE-2 versus VPNN-AE-4. This is due to training data having no anomalies of data that real-time can exhibit. While VHNN-AE-2 with single-correlation works better when testing against ground truth data, VPNN-AE-4 with double-correlation works better in real-time as the patient may not follow the Tai-Chi movements correctly. This causes worse ground truth error as the added complexity of the architecture increases noise during output, but enables better patient-error tolerance. Because of this additional noise produced of VHNN-AE-4, improvements such as larger training datasets, more sophisticated pre- and post-processing of data, as well as improvements in NN architecture will be investigated.}
 
 The neural network addressed in our preliminary work is single-modality, which generates a single type of kinematic output, either joints' location $\langle x, y, z\rangle$ or joints' rotation in Euler angle $\langle \alpha, \beta, \gamma \rangle$. There might be potential issues for single-modality network. As illustrated in our preliminary work \cite{Liang2018-Elsevier-SH, Liang2018-CHASE}, divergence is commonly observed in an XYZ-oriented network; Euler-angle-oriented network suffer from $0-360 degree$ issue \cite{Liang2018-Elsevier-SH}. A multi-modality neural network will be studied for intelligent choreography generation.
 

\begin{figure}[htp]
 	\centering
    \includegraphics[width=0.4\textwidth]{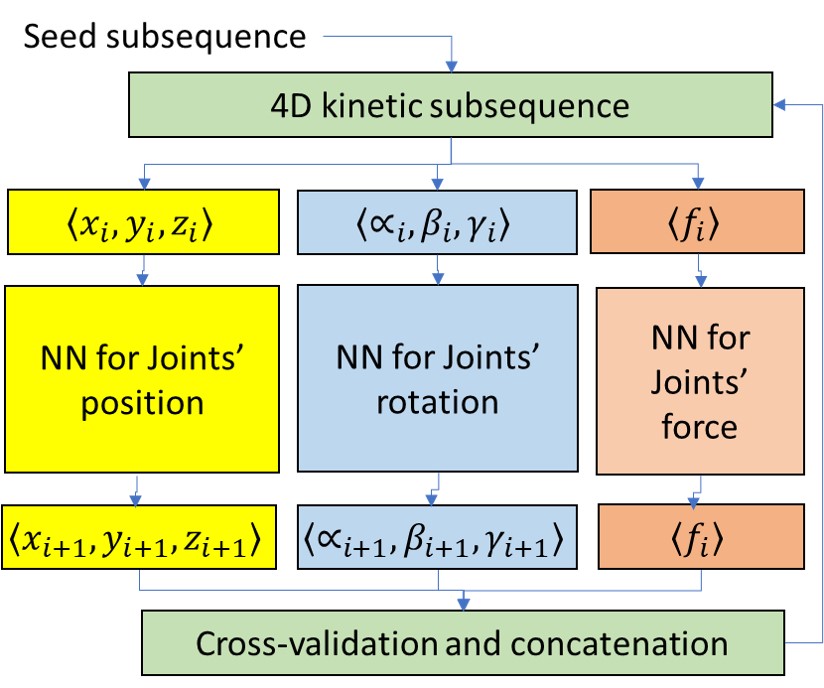}
 	\caption{Choreography using multi-modality network}
 	\label{Fig25-Choreography-MultiModality}
\end{figure} 

 As illustrated in Figure \ref{Fig25-Choreography-MultiModality}, multiple heterogeneous kinetic data will be generated simultaneously as the output of neural network so that a {\em cross-validation} can be made.

\section{Conclusion and Future Work}
\label{section:conclusion}
As seen from the reported results, the proposed multi-correlation visible hierarchical autoendoder such as VHNN-4 exhibits much better realtime generated data in comparison to MLP or classical autoencoders. Thus, the segmentation of joint data through human musculoskeletal network as well as non-divergent motion are used as effective methods to improve neural network optimization and prediction quality. In addition, the proposed work is able to generate phantom limbs with very scarce amounts of data, in contrast to previous models. Given the flexible nature of parallelized autoencoders, adding pipelines as well as the modification of individual pipelines for other applications may be easily executed.

\label{section:future work}
In the future, Generative Adversarial Networks (GAN) may be assessed for the prediction of longer time-horizons in comparison to the short time-horizons predicted by our proposed autoencoder. In addition, more pre-processing and post-processing methods may be assessed, such as singular value decomposition (SVD), in order to further aid accuracy and performance. For improvements in data acquisition, better depth sensors relative to the Microsoft Kinect V2 may be assessed for less noisy, precise data.

Future work also includes evaluating performance and user experience. A variety of user experience data and clinical benefit data will be generated during system assessment. Analytical results about the extent to which the addressed work demonstrates superior physical and psychological benefits and user experience over those of existing systems will also be derived 

\section*{Acknowledgments}
  \label{sec:Acknowledgments}
{This work is jointly sponsored by the National Science Foundation (award No: 1240734), UTC THEC/CEACSE 2016 Grant Program, and 2015 UTC CRISP program. Special thanks also go to Mr. Mr. Samuel Clark for his contributions in the paper.}

\section*{References}

\bibliographystyle{elsarticle-num}

\bibliography{reference}

\begin{thebibliography}{10}
\expandafter\ifx\csname url\endcsname\relax
  \def\url#1{\texttt{#1}}\fi
\expandafter\ifx\csname urlprefix\endcsname\relax\def\urlprefix{URL }\fi
\expandafter\ifx\csname href\endcsname\relax
  \def\href#1#2{#2} \def\path#1{#1}\fi

\bibitem{Goodfellow2016}
I.~Goodfellow, Y.~Bengio, A.~Courville, Deep Learning, MIT Press, 2016.

\bibitem{Ortiz-2016-Phantom-Pain}
M.~Ortiz-Catalan, R.~A. Guomundsdottir, M.~B. Kristoffersen,
  A.~Zepeda-Echavarria, K.~Caine-Winterberger, K.~Kulbacka-Ortiz,
  C.~Widehammar, K.~Eriksson, A.~Stockselius, C.~Ragnö, Z.~Pihlar, H.~Burger,
  L.~Hermansson, Phantom motor execution facilitated by machine learning and
  augmented reality as treatment for phantom limb pain: a single group,
  clinical trial in patients with chronic intractable phantom limb pain, The
  Lancet 388~(10062) (2016) 2885--2894.
\newblock \href {http://dx.doi.org/10.1016/S0140-6736(16)31598-7}
  {\path{doi:10.1016/S0140-6736(16)31598-7}}.

\bibitem{Ramachandran1998-PhantomLimb}
V.~S. Ramachandran, W.~Hirstein,
  \href{https://dx.doi.org/10.1093/brain/121.9.1603}{{The perception of phantom
  limbs. The D. O. Hebb lecture.}}, Brain 121~(9) (1998) 1603--1630.
\newblock \href
  {http://arxiv.org/abs/http://oup.prod.sis.lan/brain/article-pdf/121/9/1603/17864034/1211603.pdf}
  {\path{arXiv:http://oup.prod.sis.lan/brain/article-pdf/121/9/1603/17864034/1211603.pdf}},
  \href {http://dx.doi.org/10.1093/brain/121.9.1603}
  {\path{doi:10.1093/brain/121.9.1603}}.
\newline\urlprefix\url{https://dx.doi.org/10.1093/brain/121.9.1603}

\bibitem{Liang2018-Elsevier-SH}
Y.~Liang, D.~Wu, D.~Dakila, C.~Davis, R.~Slaughter, Z.~B. Guo, Virtual tai-chi
  system, a smart connected modality for rehabilitation, Elsevier: Smart Health
  9--10 (2018) 232--249.

\bibitem{online13-Dakila-Autoencoder-PhantomLeg2018}
D.~Ledesma, Y.~Liang, [online] vtcs: Construct phantom legs according to arms'
  movement using varius autoencoder strategies,
  \url{https://youtu.be/ZmTobUjRX64}, cite November 08, 2018.

\bibitem{Bengio2009}
Y.~Bengio, Learning deep architectures for ai, fundamental Trends Machine
  Learning 2~(1) (2009) 1--127.

\bibitem{Sperduti1997}
A.~Sperduti, A.~Starita, Supervised neural networks for the classification of
  structures, IEEE Transactions on Neural Networks 8~(3) (1995) 714--735.
\newblock \href {http://dx.doi.org/10.1109/72.572108}
  {\path{doi:10.1109/72.572108}}.

\bibitem{Liang2018-CHASE}
Y.~Liang, D.~Wu, D.~Dakila, C.~Davis, R.~Slaughter, Z.~B. Guo, Virtual taiji
  system, a smart-connected modality for rehabilitation, in: The Third IEEE/ACM
  Conference on Connected Health: Applications, Systems, and System
  Technologies, Washington, D.C., 2018.

\bibitem{Murphy2018-MusculoskeletanNetwork}
A.~C. Murphy, S.~F. Muldoon, D.~Baker, A.~Lastowka, B.~Bennett, M.~Yang, D.~S.
  Bassett, Structure, function, and control of the human musculoskeletal
  network. plos biology 16~(1).
\newblock \href {http://dx.doi.org/10.1371/journal.pbio.2002811}
  {\path{doi:10.1371/journal.pbio.2002811}}.

\bibitem{Farebrother2003}
R.~W. Farebrother, J.~Grob, S.-O. Troschke, Matrix representation of
  quaternions, Linear Algebra and its Applications 362 (2003) 251--255.
\newblock \href {http://dx.doi.org/doi:10.1016/s0024-3795(02)00535-9}
  {\path{doi:doi:10.1016/s0024-3795(02)00535-9}}.

\bibitem{Grood1983}
G.~ES, S.~WJ, A joint coordinate system for the clinical description of
  three-dimensional motions: application to the knee, J Biomech Eng. 105~(2)
  (1983) 136--44.

\bibitem{Shoemake1985}
K.~Shoemake, Animating rotation with quaternion curve, SIGGRAPH '85
  proceedings, Computer Graphics 19~(3) (1985) 245--255.

\bibitem{online8}
Y.~Liang, D.~Ledesma, D.~Wu, [online] vtcs: virtuality coaching based on unity
  3d, \url{https://www.youtube.com/watch?v=OQySt2i8dzo\&feature=youtu.be},
  cited May 8, 2018.

\bibitem{Tiwana2012}
M.~Tiwana, S.~Redmond, N.~Lovell, A review of tactile sensing technologies with
  applications in biomedical engineering, Sensors and Actuators A: Physical 179
  (2012) 17--31.
\newblock \href {http://dx.doi.org/10.1016/j.sna.2012.02.051}
  {\path{doi:10.1016/j.sna.2012.02.051}}.

\bibitem{online12-Dakila-EdgeComputing2018}
D.~Ledesma, W.~Baker, Y.~Liang, [online] vtcs: Deployment of vtcs over
  commodity hardware, \url{https://youtu.be/\_bB8K627JZk}, cite July 04, 2018.

\bibitem{Vasan2017-Prosthesis}
G.~{Vasan}, P.~M. {Pilarski}, Learning from demonstration: Teaching a
  myoelectric prosthesis with an intact limb via reinforcement learning, in:
  2017 International Conference on Rehabilitation Robotics (ICORR), 2017, pp.
  1457--1464.
\newblock \href {http://dx.doi.org/10.1109/ICORR.2017.8009453}
  {\path{doi:10.1109/ICORR.2017.8009453}}.

\bibitem{Lee2008}
E.~Lee, Cyber physical systems: Design challenges, in: Proc. of IEEE
  International Symposium on Object Oriented Real-time Distributed Computing
  (ISORC), 2008.

\bibitem{Sha2008}
L.~Sha, S.~Gopalakrishnan, X.~Liu, Q.~Wang, Cyber-physical systems: A new
  frontier, in: Proc. of IEEE International Conference on Sensor Networks,
  Ubiquitous and Trustworthy Computing, 2008.

\bibitem{Hu2015}
Y.~Hu, M.~Patel, D.~Sabella, N.~Sprecher, V.~Young, Mobile edge computing a key
  technology towards 5g, European Telecommunications Standards Institute (ETSI)
  White Paper.

\bibitem{Mkinen2015}
O.~Mkinen, Streaming at the edge: Local service concepts utilizing mobile edge
  computing, in: The 9th International Conference on Next Generation Mobile
  Applications, Services and Technologies, 2015.

\bibitem{Langfelder2008-HierarchicalClustering}
P.~Langfelder, S.~Horvath, Wgcna: an r package for weighted correlation network
  analysis, BMC Bioinformatics 9 (2008) 1471--2105.
\newblock \href {http://dx.doi.org/10.1186/1471-2105-9-559}
  {\path{doi:10.1186/1471-2105-9-559}}.

\bibitem{Alharbi2017}
N.~Alharbi, Y.~Liang, D.~Wu, Extended-kalman-filter preprocessing technique for
  gesture recognition, in: 2nd IEEE/ACM CHASE, Philadelphia, USA, 2017.

\bibitem{D.Ledesma2018}
D.~Ledesma, D.~Ledesma, R.~Slaughter, D.~Wu, Z.~Guo, Y.~Liang, Kinetic gesture
  recognition and choreography, in: IEEE BigDataService 2018, Bamberg, Germany,
  2018.

\bibitem{Liang2015}
Y.~Liang, Z.~B. Guo, D.~L. Wu, N.~Fell, A.~Clark, Virtual taiji system - an
  innovative modality for rehabilitation, in: Annual BSEC Conference at Oak
  Ridge National Laboratory Collaborative Biomedical Innovations, 2015.

\bibitem{C.Davis2018}
C.~Davis, D.~Ledesma, R.~Slaughter, D.~Wu, Z.~Guo, Y.~Liang, Kinetic data
  processing for gesture recognition, in: IEEE BigDataService 2018, Bamberg,
  Germany, 2018.

\bibitem{online1}
Y.~Liang, D.~Ledesma, C.~Davis, R.~Slaughter, D.~Wu, [online] vtcs:
  preprocessed and raw kinematic data (front view),
  \url{https://www.youtube.com/watch?v=KFBV\_EBTEQw}, cited March 2, 2018.

\bibitem{online2}
Y.~Liang, D.~Ledesma, C.~Davis, R.~Slaughter, D.~Wu, [online] vtcs:
  preprocessed and raw kinematic data (front and back view),
  \url{https://youtu.be/Ee31SdhXxXc}, cited March 2, 2018.

\bibitem{Ambrosio2007}
J.~Ambrosio, A.~Kecskemethy, Multibody dynamics of biomechanical models for
  human motion via optimization, Multibody Dynamics: Computational Method and
  Application, J.C. Garcia Orden et al. (Eds) 6 (2007) 245--272.
\newblock \href {http://dx.doi.org/10.1007/978-1-4020-5684-0-12}
  {\path{doi:10.1007/978-1-4020-5684-0-12}}.

\bibitem{Eich-Soellner2008}
E.~Eich-Soellner, C.~Führer, Numerical Methods in Multibody Dynamics, Teubner,
  Stuttgart, 2008.

\bibitem{Moreira2013}
P.~Moreira, U.~Lugrís, J.~Cuadrado, P.~Flores,
  \href{http://hdl.handle.net/1822/23057}{Biomechanical models for human gait
  analyses using inverse dynamics formulation}, Sociedade Portuguesa de
  Biomecânica, 2013.
\newline\urlprefix\url{http://hdl.handle.net/1822/23057}

\bibitem{Liang2013-FEM}
Y.~Liang, M.~Szularz, L.~T. Yang, Finite-element-wise domain decomposition
  iterative solvers based on polynomial preconditioning, Mathematical and
  Computer Modeling 58~(1-2) (2013) 421--437.
\newblock \href {http://dx.doi.org/10.1016/j.mcm.2012.11.017}
  {\path{doi:10.1016/j.mcm.2012.11.017}}.

\bibitem{Wittenburg2008}
J.~Wittenburg, Dynamics of Multibody Systems, Berlin, Springer, 2008.

\bibitem{Kalman1960}
R.~Kalman, A new approach to linear filtering and prediction problems, Journal
  of Basic Engineering 80 (1960) 35.
\newblock \href {http://dx.doi.org/doi:10.1115/1.3662552}
  {\path{doi:doi:10.1115/1.3662552}}.

\bibitem{Einicke2012}
G.~Einicke, Smoothing, Filtering and Prediction: Estimating the Past, Present
  and Future, Rijeka, Croatia: Intech, 2012.

\bibitem{Zarchan2000}
P.~Zarchan, H.~Musoff, Fundamentals of Kalman Filtering: A Practical Approach,
  American Institute of Aeronautics and Astronautics, Incorporated, 2000.

\bibitem{butepage2017-deepmotionnn}
J.~Butepage, M.~J. Black, D.~Kragic, H.~Kjellstrom, Deep representation
  learning for human motion prediction and classification, in: Proceedings of
  the IEEE Conference on Computer Vision and Pattern Recognition, 2017, pp.
  6158--6166.

\bibitem{online20-Dakila-PhantomLeg2018}
D.~Ledesma, Y.~Liang, [online] playlist of phantom legs induced by moving arms,
  \url{https://www.youtube.com/playlist?list=PL8VO3nxhbh0IngDo5ij0PkKxK5uFbtXE0},
  cite December 08, 2018.

\bibitem{online14-Dakila-MLP-PhantomLeg2018}
D.~Ledesma, Y.~Liang, [online] vtcs: Construct phantom legs according to arms'
  movement using 7-layer mlp, \url{https://youtu.be/BFW3JVI0dKM}, cite December
  08, 2018.

\bibitem{qiao2019-hierchicalcnnwords}
X.~Qiao, C.~Peng, Z.~Liu, Y.~Hu, Word-character attention model for chinese
  text classification, International Journal of Machine Learning and
  Cybernetics (2019) 1--17.

\bibitem{Li2015-hierchicalaewords}
J.~Li, M.-T. Luong, D.~Jurafsky, A hierarchical neural autoencoder for
  paragraphs and documents, arXiv preprint arXiv:1506.01057.

\bibitem{Yian2019-hierchicalcnnfashion}
Y.~Seo, K.~Shin, Hierarchical convolutional neural networks for fashion image
  classification, Expert Systems with Applications 116 (2019) 328--339.

\bibitem{online15-Dakila-MLP-PhantomLeg2018}
D.~Ledesma, Y.~Liang, [online] vtcs: Construct phantom legs according to arms'
  movement using autoencoder, \url{https://youtu.be/1aHJQyST9k0}, cite December
  08, 2018.

\bibitem{online16-Dakila-MLP-PhantomLeg2018}
D.~Ledesma, Y.~Liang, [online] vtcs: Construct phantom legs according to arms'
  movement using two-thead vhnn autoencoder,
  \url{https://youtu.be/pVtzw0dJduA}, cite December 08, 2018.

\bibitem{online17-Dakila-MLP-PhantomLeg2018}
D.~Ledesma, Y.~Liang, [online] vtcs: Construct phantom legs according to arms'
  movement using four-thead vhnn autoencoder,
  \url{https://youtu.be/dGPm4z7vibI}, cite December 08, 2018.

\end{thebibliography}





\end{document}